\documentclass[aps,prd,amsfonts,amsmath,nofootinbib,showpacs,preprint,tightenlines,longbibliography]{revtex4-1} 
\usepackage{graphicx}

\def\be{\begin{equation}}
\def\ee{\end{equation}}
\def\bea{\begin{eqnarray}}
\def\eea{\end{eqnarray}}
\def\bml{\begin{subequations}}

\def\elea{\end{eqnarray}\end{subequations}}

\def\bcrit{\beta_{\text{crit}}}

\begin{document}

% This works around a bug RevTex4.1 in 1-column format.
\count\footins = 1000

\title{Energy-conservation constraints on cosmic string loop
  production and distribution functions}

\author{Jose J. Blanco-Pillado}
\email{josejuan.blanco@ehu.es}
\affiliation{IKERBASQUE, Basque Foundation for Science, 48011, Bilbao, Spain}
\affiliation{Department of Theoretical Physics, UPV/EHU,\\48080, Bilbao, Spain}

\author{Ken D. Olum}
\email{kdo@cosmos.phy.tufts.edu}
\affiliation{Institute of Cosmology, Department of Physics and Astronomy,\\Tufts University, Medford, MA 02155, USA}

\author{Jeremy M. Wachter}
\email{jeremy.wachter@ehu.es}
\affiliation{Department of Theoretical Physics, UPV/EHU,\\48080, Bilbao, Spain}

\begin{abstract}
A network of cosmic strings would lead to gravitational waves which
may be detected by pulsar timing or future interferometers.  The
details of the gravitational wave signal depend on the distribution
of cosmic string loops, which are produced by intercommutations from
the scaling network of long strings.  We analyze the limits imposed by
energy conservation, i.e., by the fact that the total amount of string
flowing into loops cannot exceed the amount leaving the long strings.
We show that some recent suggestions for the cosmic string loop
production rate and distribution are ruled out by these limits.  As a
result, gravitational waves based on such suggestions, in particular
``model 3'' used in LIGO data analysis, are not to be expected.
\end{abstract}

\maketitle

\section{Introduction}

The universe may contain a network of cosmic strings formed at a
symmetry breaking transition in the early universe or by brane
inflation in string theory.  (For reviews see
\cite{Vilenkin:2000jqa,Polchinski:2004ia}.)  In the simplest cases,
which we will discuss here, strings have neither ends nor vertices, so
the network (in a spatially infinite universe) consists of infinite
strings and closed loops.

The strings are continually intercommuting, so that loops may
break off of infinite strings or rejoin to them, and loops may also
fragment or join with each other.  However, the net effect is a
production of loops, so that string energy leaves the infinite string
network and flows into the loop distribution.  Loops then oscillate,
emitting gravitational waves and eventually decaying.  These processes
allow for the cosmic string network to reach a scaling regime, in
which all linear measures evolve (on average) in a way proportional to
the cosmic time $t$.  The energy density of the scaling network
evolves as radiation in the radiation era and matter in the matter
era, so that the string network is always a subdominant component and
does not cause the problems that monopoles would.

Gravitational waves are the leading way to look for a cosmic string
network
\cite{Sanidas:2012ee,Binetruy:2012ze,Kuroyanagi:2012wm,Sousa:2016ggw,Blanco-Pillado:2017rnf,Ringeval:2017eww,Abbott:2017mem,Arzoumanian:2018saf,Burke-Spolaor:2018bvk,Siemens:2019xkk}.
The observable gravitational waves come mostly from the loop
distribution, and therefore it is of great importance to understand
this distribution.  Many distributions have been inferred from
simulations or proposed on theoretical grounds.  Here we discuss some
important constraints on the rate of loop production and the resulting
distribution of loops, arising from the fact that the energy in loops
comes from energy originally in the long string network, so that
energy conservation couples the loop production rate to the loss of
energy in the long strings.

In the next section we discuss the definition of the long string
network and the loop production and distribution functions.  In
Sec.~\ref{sec:conservation}, we derive and apply the constraints
resulting from energy conservation in the production of loops and
compare with specific models of loop production.  In
Sec.~\ref{sec:no-smoothing}, we point out that the problem is more
general than a conflict of numerical values but applies to any attempt
to derive a rapidly-diverging loop production functions from a
simulation, and in Sec.~\ref{sec:loopdistribution} we point out that
these constraints apply also to certain loop distribution functions.
We conclude in Sec.~\ref{sec:conclusion}.

A dictionary for translating between the parameters and functions used
here and those in some other papers is given in Appendix~\ref{sec:notation}.

\section{Loops and long strings}\label{sec:classification}

The separation between loops and infinite strings is not
completely straightforward, because a sufficiently long loop cannot be
cleanly distinguished from an infinite string.  Loops much larger than
the horizon are continually reconnecting to infinite strings and
breaking off from them again.  A very long loop consists of many
causally disconnected segments, and the dynamics of each segment may
connect the loop to an infinite string \cite{Kibble:1985tf-fixed}, so the
typical lifetime of such a loop between intercommutations drops
inversely with the loop length.

Loops much smaller than the horizon, however, are very unlikely to
join with other strings, because these loops are much smaller than the
distance between strings, which grows with the expansion of the
universe.  Small loops, once formed, may fragment into smaller loops,
but simulations show that this process does not continue indefinitely
but rather yields a distribution of non-self-intersecting
loops.\footnote{Fragmentation is less common than one might at first
  think.  Loops are formed by the combination of right-moving and
  left-moving excitations on the string.  When a loop forms, many
  small excitations have already passed through each other without
  forming loops, so they will not do so on future oscillations.
  Others have not yet passed, and may thus form a smaller loop in the
  first oscillation, but this loop forms with no causal dependence on
  the fact that it is part of a larger loop at the time of formation.
  The only small loops that form because of being on a larger loop are
  those which include at least one of the kinks arising from the
  larger loop's formation \cite{Thompson:1988yj}.  Simulations show
  this phenomenon to be quite rare.} It is thus possible to make a
reasonably clear distinction between loops, meaning small loops on
non-self-intersecting trajectories and that we do not expect to rejoin
larger structures, and long strings, in which we include both
super-horizon loops and strings that really are infinite.

Simulations, of course, have no infinite strings.  Simulators
generally use periodic boundary conditions, meaning that all strings
are in loops.  Typically all strings that cross the horizon are part
of a single large loop that crosses through the periodic boundary
conditions many times.  Again it is possible to distinguish small
loops from long strings, meaning loops above a certain size.  In our
simulations \cite{BlancoPillado:2011dq,Blanco-Pillado:2013qja}, we
define loops existing at a certain time as closed strings of any
length that will not self-intersect or rejoin in the
future,\footnote{After a loop has oscillated three times, we remove it
  from the simulation, so we would miss rejoinings after that stage.
  But we have experimented with allowing many more oscillations before
  removal, and this makes no significant difference to any quantity
  reported by the simulation.  We run long enough beyond the reported
  simulation ending time to allow loops of up to half the horizon size
  to undergo the necessary number of oscillations to be correctly
  classified as loops.} but the exact definition will not be
important here, especially as we will mainly be concerned with loops
far below the horizon size.

We will describe loops at time $t$ by a loop distribution function, 
$n(l,t)$, that gives the density of loops per unit volume per unit loop 
length existing at time $t$.  We will describe loop production by a function $f(l,t)$ giving
the number of loops produced per unit time per unit volume per unit
loop length.  Loops in self-intersecting trajectories are
excluded from both of these functions.  We also exclude loops that will join
to long strings or other loops, but this is of little consequence for
loops much smaller than the typical interstring distance, because it
is very unlikely that they will find any other string to join.

All lengths here are invariant, i.e., a loop of length $l$ has energy
$\mu l$, where $\mu$ is the energy per unit length (tension) of the string, 
and we work in units where the speed of light is set to 1.  The energy 
density in long strings (i.e., everything that is not counted in $n(l,t)$) will be denoted
$\rho_\infty$.

\section{Energy conservation}\label{sec:conservation}

The breaking off of loops conserves energy, so that the total
invariant length of string before and after an intercommutation is the
same.\footnote{We neglect a tiny amount of particle radiation here.
  Taking account of it would only strengthen our conclusions.}  This
leads to a constraint \cite{Kibble:1984hp-fixed}, because the energy
flowing into loops must flow out of long strings.  The long-string
energy density also decreases due to dilution of strings and
redshifting of the string velocity due to the expansion of the
universe.  The resulting evolution equation for the energy density of long
strings is 
\be\label{eqn:drdt}
\frac{d\rho_\infty}{dt} = -2 H \left(1 + \left<v^2_\infty\right>
\right)\rho_\infty - \mu \int_0^\infty l f(l,t)\,dl\,,
\ee
where $H$ is the Hubble constant and $\left<v^2_\infty\right>$ is the
rms average velocity of the long strings.  Equation~(\ref{eqn:drdt})
constrains the total rate of loop production.

In this paper, the parameter $l$ refers to the invariant length of the
loop at the time of production, i.e., its total energy divided by
$\mu$.  Some of this energy is in the overall kinetic energy of the
loop (with respect to the Hubble flow).  If the loop is long-lived
compared to the Hubble time, this kinetic energy will be lost to
redshifting, so what matters is the rest energy
\cite{Blanco-Pillado:2013qja}.  For very short loops, which will be of
most concern to us here, $l$ is the natural variable.\footnote{If we
  let $m$ be loop rest energy, the total rest mass appearing in loops
  is $\int_0^\infty m f(m,t)\,dm$, which is less than $\mu
  \int_0^\infty l f(l,t)\,dl$, leading to a stronger constraint on
  $f(m,t)$ than on $f(l,t)$.}

No model of the string network is necessary for Eq.~(\ref{eqn:drdt}),
but we can go further if we assume that the network is in a scaling
regime in a cosmological era where the scale factor $a\propto t^\nu$ so
that $\nu = 1/2$ in the radiation era and $2/3$ in the matter era.  In
that case we define a scaling measure of the loop length, $x = l/t$,
and define $n(x) = t^4n(l, t)$ to be the number of loops per unit $x$
in volume $t^3$, $f(x) = t^5f(l, t)$ to be the number of loops per
unit $x$ produced in time $t$ in volume $t^3$, and the ``interstring
distance'' $\gamma = \sqrt{\mu/\rho_\infty}/t$.  In a scaling regime,
$\gamma$ is constant, and $n(x)$ and $f(x)$ depend only on $x$ and
not on $t$.  In that description, Eq.~(\ref{eqn:drdt}) becomes
\be\label{eqn:xf}
\int_0^\infty x f(x)\,dx = 
\frac{2}{\gamma^2}
\left(1 - \nu(1+\langle v^2_\infty\rangle)\right)\equiv B\,.
\ee

Any proposed scaling loop production function $f(x)$ must obey
Eq.~(\ref{eqn:xf}).  In our simulations\footnote{The definition of
  loops and loop production used in our simulations is exactly as
  described above.  However, it would be difficult to report long
  string statistics in a way which depends on the future evolution of
  the string.  Instead we report our $\rho_\infty$ and $\langle
  v^2_\infty\rangle$ including all string that has not been identified
  as being in non-self-intersecting loops.  Some string in loops may
  later rejoin, and some string is in loops that we have not yet
  identified.  However, we can recognize both of these phenomena
  later.  The maximum error they could have introduced is less than
  1\%.}~\cite{BlancoPillado:2011dq} (values from other groups are very
similar), we find $\gamma = 0.30$ and $\langle v^2_\infty\rangle=
0.40$ in the radiation era, and $\gamma = 0.51$ and $\langle
v^2_\infty\rangle= 0.35$ in the matter era,
so\footnote{Reference~\cite{BlancoPillado:2011dq} defined scaling
  quantities in terms of the horizon distance and consequently the $B$
  found there was larger by factor $(1-\nu)^{-3}$.  See
  Appendix~\ref{sec:notation}.}
\be\label{eqn:bounds}
B \approx \begin{cases}6.7 & \text{radiation,}\\
0.77& \text{matter.}\end{cases}
\ee

Reference~\cite{Lorenz:2010sm} discusses loop production functions
(further analyzed as part of Ref.~\cite{Auclair:2019zoz}) which grow
rapidly toward small scales until they are cut off at some value
$x_c$, which is intended to represent the effect of gravitational
smoothing on long strings.  Specifically, they consider the
possibility that\footnote{Equation (2.15) of
  Ref.~\cite{Auclair:2019zoz} includes a second term representing
  reduced but nonzero production of loops at scales below $x_c$.
  Including it would increase the energy flow into loops and so make
  the conflict here worse.}

\be\label{eqn:fpower}
f(x) = c x^{-\beta} \Theta(x-x_c)\,,
\ee
with $\beta >2$.  Integrating Eq.~(\ref{eqn:fpower}) gives
\be\label{eqn:ftotal}
\int_0^\infty x f(x)\,dx = \frac{c}{(\beta-2)x_c^{\beta-2}}\,.
\ee

Following Ref.~\cite{gr-qc/0702055},
Refs.~\cite{Lorenz:2010sm,Auclair:2019zoz} say that we should take
\be\label{eqn:xc}
x_c = \Upsilon (G\mu)^{4-\beta}\,,
\ee
with $\Upsilon\sim20$, and suggest that we choose $\beta$ and $c$ to
match the results of Ref.~\cite{Ringeval:2005kr}.  Thus $\beta = 2.6$
in the radiation era and 2.4 in the matter era.\footnote{These are the
  central values given in Ref.~\cite{Ringeval:2005kr}.  For other
  possibilities see Appendix~\ref{sec:errorbars}.}  With $G\mu= 10^{-7}$
as suggested by Ref.~\cite{Auclair:2019zoz}, Eq.~(\ref{eqn:xc}) gives
$x_c\approx 3\times 10^{-9}$ (radiation), $1\times 10^{-10}$ (matter).
To find a corresponding value of $c$ we use Eq.~(2.22) of
Ref.~\cite{Auclair:2019zoz},\footnote{Since we are matching
  Ref.~\cite{Ringeval:2005kr}, where there is no gravitational
  backreaction, we do not need the more general form of Eq.~(2.17) of
  Ref.~\cite{Auclair:2019zoz}.} which in our notation gives
\be\label{eqn:nA}
n(x) =\frac{c}{\beta-\bcrit}x^{-\beta}\,,
\ee
where $\bcrit = 4-3\nu = 5/2$ (radiation), 2 (matter).
Our $n(x)$ corresponds to
\be
(1-\nu)^4\frac{S(\alpha)}{\alpha} =(1-\nu)^4C_0\alpha^{-p-1}
\ee
in the notation of Ref.~\cite{Ringeval:2005kr}, where $C_0 =
0.21$ (radiation), 0.09 (matter). In our notation,
\be\label{eqn:nRSB}
n(x)=(1-\nu)^{4-\beta}C_0 x^{-\beta}\,.
\ee
Setting Eqs.~(\ref{eqn:nA}) and (\ref{eqn:nRSB})
equal gives
\be\label{eqn:cC0}
c = (\beta-\bcrit) C_0 (1-\nu)^{4-\beta}
\ee
and the numbers above give $c= 0.008$ in the radiation era and 0.006 in
the matter era.

Putting these values in Eq.~(\ref{eqn:ftotal}) gives
\be
\int_0^\infty x f(x)\,dx \approx 
\begin{cases}2000 & \text{radiation,}\\
200& \text{matter.}\end{cases}
\ee
larger than the values in Eq.~(\ref{eqn:bounds}) by a factor more than
200 in both cases.  In fact the situation is much worse than that,
because non-observation of gravitational waves limits $G\mu$ to be no
more than of order $10^{-11}$
\cite{Lentati:2015qwp,Blanco-Pillado:2017rnf,Abbott:2017mem,Arzoumanian:2018saf},
and then the discrepancy is $6\times 10^5$ in the radiation era and
$9\times 10^4$ in the matter era.

Thus it is impossible to have the loop production function of
Eq.~(\ref{eqn:fpower}) with parameters at all similar to those used by
Refs.~\cite{Ringeval:2005kr,Lorenz:2010sm} and discussed in
Ref.~\cite{Auclair:2019zoz}.

\section{Networks without gravitational smoothing}\label{sec:no-smoothing}

Suppose we had a huge computer and could run large simulations
(without gravitational effects) for as long as we wished.  We could
continue deep into the scaling regime,\footnote{Without gravitational
  effects we cannot have true scaling, because energy will collect in
  tiny loops.  But we would expect scaling in the long string network
  and in loops above some continually decreasing lower limit size.}
and discover the scaling loop production function.  What could it be?
Suppose it were a power law for small $x$.  There would be no
gravitational cutoff.  So we would have just $f(x) = c x^{-\beta}$ and
if $\beta>2$, $\int x f(x)\, dx$ would diverge at $x=0$.  Such a
scenario would violate \emph{any} energy conservation bound, so no
such result is possible.  We cannot evade this conclusion by proposing
cutoffs due to gravitational scales, because the hypothesized
simulation does not include gravitation.

Our hypothetical simulation would have to give some $f(x)$ that obeyed
energy conservation and so did not rise too quickly at small scales.
It would give an $n(x)$ that would go as $x^{-\bcrit}$ at small $x$.
If we had such a simulation, we could then apply gravitational effects
to give an updated $n(x)$, which would then go as $x^{-\bcrit}$ down
to some $x$ where gravitational effects became important, and then
fall below that line.

But we do not have such a simulation.  Instead we must make do by
extrapolation from smaller simulations.  In
Ref.~\cite{Ringeval:2005kr}, the authors fit $n(x)$ to a power law
over the range where they felt $n(x)$ was accurately determined.  What
can we do with this information?  It would be an error to extrapolate
this power law and conclude that in a much larger simulation it would
continue forever, because we know that is impossible.  Neither would
it make sense to extrapolate the power law until some gravitational
cutoff, because the simulation does not include gravitation.  We know
that in the nongravitational world, there will eventually be some new
behavior, but we don't know what it is.  Since it is wrong to
extrapolate the power law form of $n(x)$ in the simulated world, it
would be wrong to extrapolate in the real world.  Thus it does not
make sense to use the power law $n(x)$ from a simulation to derive
$n(x)$ in the real universe at any smaller $x$ than those where the
simulation finds scaling behavior.

\section{Constraints on the loop distribution}\label{sec:loopdistribution}

The argument above constrains not only the loop production function
but also the loop distribution.  The number of loops can be found by
integrating the production function, accounting for the decrease in
loop size due to gravitational backreaction.  In a scaling regime
\cite{Blanco-Pillado:2013qja},
\be
n(x) = \frac{\int_x^\infty (x'+\Gamma G \mu)^{\bcrit-1} f(x') dx'}{(x+\Gamma G \mu)^{\bcrit}}\,.
\ee
For $x\gg\Gamma G \mu$,
\be\label{eqn:nnoG}
n(x) = \frac{\int_x^\infty x'^{\bcrit-1} f(x') dx'}{x^{\bcrit}}\,.
\ee
If the integral in the numerator does not depend on the lower
limit as $x\to0$, we find $n(x)\sim x^{-\bcrit}$.  If $n(x)$ diverges more
rapidly than this as $x$ decreases, the divergence must come partly
from the numerator.  The only way to have $n(x)\sim x^{-\beta}$ with
$\beta >\bcrit$ is to have $f(x)\sim x^{-\beta}$. In other words, a
distribution $n(x)\sim x^{-\bcrit}$ may arise from loops produced at
earlier times, but $n(x)$ may only diverge more rapidly than this if
the tiny loops in question were produced very recently by a similarly
diverging production function.  But no argument based on simulation
could support such a production function.

More directly, any scaling loop production function must obey
Eq.~(\ref{eqn:bounds}) to conserve energy, so the integral in that
equation must converge.  In a simulation that does not include
gravitational radiation effects, the relationship between $f(x)$ and $n(x)$ is
given by Eq.~(\ref{eqn:nnoG}).  Since $\bcrit\ge 2$, the integral in
the numerator of Eq.~(\ref{eqn:nnoG}) must also converge even if $x$
is taken to 0.  Thus for small enough $x$, $n(x)\sim x^{-\bcrit}$ and
cannot diverge any faster.

Thus no simulation can find $n(x)\sim x^{-\beta}$ with $\beta >\bcrit$
for arbitrarily small $x$.  The loop distribution suggested in
Ref.~\cite{Lorenz:2010sm} with $\beta = 2.6$ in the radiation
era\footnote{Note that $n(x)\sim x^{-2.5}$ is within the error bars of
  Ref.~\cite{Ringeval:2005kr}, and if that is the true shape of
  $n(x)$, there is no problem with energy conservation.  But the
  conclusion in that case is very different.  We can get $n(x)\sim
  x^{-2.5}$ from a wide range of loop production functions, even a
  $\delta$-function~\cite{Vilenkin:2000jqa,Blanco-Pillado:2013qja,Auclair:2019zoz}.}
cannot be supported by the simulations of Ref.~\cite{Ringeval:2005kr}.
Therefore there is no reason to use gravitational-wave predictions
based on this spectrum, in particular ``model 3'' of
Ref.~\cite{Abbott:2017mem}.

This criticism does not affect ``model 1'' and ``model 2'' of
Ref.~\cite{Abbott:2017mem}.  Both of these models involve a loop
production function which is peaked at a certain range of scales not
depending on any gravitational cutoff. In such a model,
Eq.~(\ref{eqn:ftotal}) gives some finite number independent of $x_c$,
and the only issue is that that number should agree with
Eq.~(\ref{eqn:xf}).  In ``model 1'', loops are all produced at the
same scale (relative to the the age of the universe), and the
production rate is adjusted to make Eq.~(\ref{eqn:xf}) hold.  ``Model
2'' takes the loop density from Ref.~\cite{Blanco-Pillado:2013qja},
which is based on the production function found in
Ref.~\cite{BlancoPillado:2011dq}, and we checked in
Ref.~\cite{BlancoPillado:2011dq} that indeed the $f(x)$ found there
obeys Eq.~(\ref{eqn:xf}).

\section{Conclusion}\label{sec:conclusion}

To predict observable signals, such as gravitational waves, from a
cosmic string network requires knowledge of the distribution of loops
at times when the signals may be emitted.  To obtain that knowledge we
use simulations, but we cannot simulate the cosmologically necessary
range of scales, so we must extrapolate from simulations.  However, it
does not make sense to use loop production functions that do not
conserve energy, nor to use loop distributions that can result only
from such unrealistic production functions.

Of particular concern are loop production functions of the form
$cx^{-\beta}$ with $\beta > 2$.  If not cut off, such a function leads
to an infinite flow of energy into loops.  A cutoff will make the flow
finite, but the actual gravitationally-based cutoffs proposed for this
purpose yield an energy flow much larger than is available from the
scaling network of long strings.  With modern limits on $G\mu$, the
discrepancy is more than $10^5$.  This is much too large to be
explained by any effects such as small-scale structure or
field-theoretic excitations on long strings.  Thus loop production
functions of this form, and loop distribution functions arising from
them, should not be used to calculate observable effects.

\section*{Acknowledgments}

We are grateful to Ana Achucarro, Leandros Perivolaropoulos, Tanmay
Vachaspati, and the Lorentz Center for bringing together the cosmic
string community at the workshop ``Cosmic Topological Defects:
Dynamics and Multi-messenger Signatures'', and to the participants of
that workshop, especially Pierre Auclair, Christophe Ringeval, Mairi
Sakellariadou and Dani\`ele Steer.  We thank the anonymous
referee for raising the issues that became Appendix B.

This work was supported in part by the National Science Foundation
under grant number 1820902, the Spanish Ministry MINECO grant
(FPA2015-64041-C2-1P), the Basque Government grant
(IT-979-16) and the MCIU/AEI/FEDER grant (PGC2018-094626-B-C21). 
J. J. B.-P. is also supported in part by the Basque
Foundation for Science (IKERBASQUE).

\appendix
\section{Dictionary of notations}\label{sec:notation}
Different papers use different symbols to denote the same
concept, and furthermore even when the concept is the same, some
papers use the cosmic time $t$ to define scaling units, while
others use the horizon distance $d_h = t/(1-\nu)$.  As a result,
powers of $1-\nu$ are needed to convert between values given in the
different papers.  Table~\ref{tab:notation} lists the notations and
the conversion factors for this paper and several recent works.

\begin{table}
\begin{equation*}
\begin{array}{|c|c|c|c|l|}
\hline
\text{this paper} & \text{BOS \cite{BlancoPillado:2011dq}} &
\text{RSB \cite{Ringeval:2005kr}} & \text{ARSS \cite{Auclair:2019zoz}}&\\
\hline
\mu & \mu & U & U &\text{string tension}\\
x & (1-\nu)^{-1}x & (1-\nu)^{-1}\alpha & \gamma &\text{loop length}\\
\gamma & (1-\nu)^{-1}\gamma & &\gamma_\infty& \text{interstring distance}\\
f(x) & (1-\nu)^5f(x) & & t^5\mathcal{P} &\text{loop creation rate}\\
n(x) & (1-\nu)^4n(x) &(1-\nu)^4\mathcal{S(\alpha)}/\alpha& t^4\mathcal{F}&\text{loop distribution}\\
\beta& &p+1 & 3-2\chi&\text{exponent in loop creation/distribution}\\
B &(1-\nu)^3\mathcal{P}  &  & & \text{energy flow into loops}\\
\hline
\end{array}
\end{equation*}
\caption{Notations used here and in some recent papers.  The
  quantities in the different columns in each row are the same when
  evaluated using the conventions and notation of the papers listed.}
\label{tab:notation}
\end{table}

\section{Range of exponents in $n(x)$}\label{sec:errorbars}
Reference~\cite{Ringeval:2005kr} gave error bars on the possible
exponents derived from their simulation.  In our notation, \
\be\label{eqn:fullbeta}
\beta = \begin{cases}
2.60^{+0.21}_{-0.15} & \text{radiation}\\
2.41^{+0.08}_{-0.07} & \text{matter}\,.
\end{cases}
\ee
In the main text we considered only the central values of this
parameter; here we will consider whether other possibilities for the
exponent will allow these distributions to escape the bounds above.

First consider the radiation era.  The error bars above allow the
possibility that $\beta = 2.5$.  If that is correct, $n(x)$ could
arise from a wide range of distribution functions including those
discussed in Refs.~\cite{BlancoPillado:2011dq,Blanco-Pillado:2013qja}.
We cannot then include gravitational effects without knowing more
about the loop production function.  In particular, such an $n(x)$
from simulations cannot be used as evidence for a diverging loop
distribution in the real universe with gravitation.

Another possibility is that $\beta=5/2+\epsilon$ with $\epsilon\ll1$.
From Eq.~(\ref{eqn:cC0}) it appears that $c$ would be very small and
so the energy conservation bounds could be obeyed.  However, in such
a regime we should consider the loop production function more
carefully.  Without gravity, the relationship between $n(x)$ and
$f(x)$ in the radiation era is~\cite{BlancoPillado:2011dq}
\be
n(x) = x^{-5/2} \int_x^\infty x'^{3/2} f(x') dx'
\ee
in our notation.  As we discussed in Sec.~\ref{sec:classification}
above, loops larger than the horizon rarely survive and should not be
counted in $f(x)$.  Thus Eq.~(\ref{eqn:nA}) must be modified.  For
$1>x>x_c$ we have \cite{Auclair:2019zoz}
\be
n(x) =c x^{-5/2} \int_x^1 x'^{3/2-\beta} dx'
= \frac{c}{\epsilon} \left[x^{-\beta}- x^{-5/2}\right]\,.
\ee
When we match this to Eq.~(\ref{eqn:nRSB}), we find
\be\label{eqn:cnew}
c = 2^{\beta-4} \frac{\epsilon C_0}{1-x^\epsilon}\,.
\ee
This depends on the value of $x$ used to determine $c$.  To attempt to
comply with energy conservation bounds, we would like to make $c$ the
smallest possible.  Since $c$ is an increasing function of $x$, we
will use the smallest $x$ in the range used in
Ref.~\cite{Ringeval:2005kr} to determine $\beta$, which is about
$5\times 10^{-3}$.

As we decrease $\epsilon$, $c$ will decrease, but never below its
$\epsilon\to0$ limit, $-C_0/(2\sqrt{2}\ln x)$.  With $x= 5\times
10^{-3}$ and $C_0 = 0.21$, we find\footnote{The reason this is larger
  than the 0.008 that we found above is that taking account of the
  finite range of $x$ where $f(x)$ contributes to the $n(x)$ found in
  Ref.~\cite{Ringeval:2005kr} is more important than reducing
  $\epsilon$ to any (positive) value.}  $c = 0.014$.  Putting this in
Eq.~(\ref{eqn:ftotal}) with $\beta = 2.5$ gives\footnote{The upper
  limit on $x$ never makes any significant difference in
  Eq.~(\ref{eqn:ftotal}) in the radiation era.} $\int_0^\infty x
f(x)\,dx \approx 1000$ for $G\mu= 10^{-7}$, still many times larger
than the value in Eq.~(\ref{eqn:bounds}).

In the matter era, the situation is different, because the power of
$x$ that multiplies $f(x)$ in the energy flow is the same one that
multiplies $f(x)$ in the calculation of $n(x)$.  If we make $\beta$
small enough, it is indeed possible to obey the energy conservation
constraint.  However, the requisite $\beta$ is about 2.15 if we take
$G\mu = 10^{-7}$ or 2.10 if we take $G\mu = 10^{-11}$ (including only
a finite range of $x$ makes little difference here).  These $\beta$ lie far
outside the range given by Eq.~(\ref{eqn:fullbeta}).

\bibliography{paper,no-slac}

\end{document}